\documentclass[runningheads]{llncs}
\usepackage{graphicx}
\usepackage{amsmath}
\usepackage{comment}
\usepackage[ruled,linesnumbered]{algorithm2e}
\usepackage{amssymb}
\usepackage{cleveref}
\usepackage{subcaption}
\usepackage[font=small]{caption}

\begin{document}
\newcommand{\squeezeup}{\vspace{-3.0mm}}
\setlength{\textfloatsep}{5.0mm}
\title{Data-Driven Substructuring Technique for Pseudo-Dynamic Hybrid Simulation of Steel Braced Frames}

\author{
Fardad Mokhtari\orcidID{0000-0001-7385-0098} \and
Ali Imanpour
		}

\titlerunning{Data-Driven Substructuring Technique for Hybrid Simulation}
\authorrunning{F. Mokhtari, A. Imanpour}

\institute{University of Alberta, Edmonton, Alberta, Canada\\
\email{fardadmokhtari@ualberta.ca}\\
\email{imanpour@ualberta.ca} \\
			}		
\maketitle

\begin{abstract}
This paper proposes a new substructuring technique for hybrid simulation of steel braced frame structures under seismic loading in which a new machine learning-based model is used to predict the hysteretic response of steel braces. Corroborating numerical data is used to train the model, referred to as PI-SINDy, developed with the aid of the Prandtl-Ishlinskii hysteresis model and sparse identification algorithm. By replacing a brace part of a prototype steel buckling-restrained braced frame with the trained PI-SINDy model, a new simulation technique referred to as data-driven hybrid simulation (DDHS) is established. The accuracy of DDHS is evaluated using the nonlinear response history analysis of the prototype frame subjected to an earthquake ground motion. Compared to a baseline pure numerical model, the results show that the proposed model can accurately predict the hysteretic response of steel buckling-restrained braces.


\end{abstract}

\section{Introduction}

Hybrid simulation is an advanced structural testing method in which the critical component of the structure expected to experience instability or significant nonlinearity is tested physically in the laboratory while the rest of the structure is numerically analyzed using a parallel finite element model \cite{mccrum2016overview}. In earthquake engineering, there are generally two approaches to perform hybrid simulation: 1) pseudo-dynamic hybrid simulation (PsDHS), where the simulation is conducted at a slower pace compared to the time of ground motion acceleration, and 2) real-time hybrid simulation (RTHS) where the simulation is performed in real-time, offering a favorable option for structures sensitive to strain rate effects.

Hybrid simulation offers a viable and economical response evaluation method over the conventional quasi-static and shaking table testing methods. However, the results of hybrid simulation may become biased when only one or a limited number of critical elements  (e.g., seismic fuses) in the structure is physically tested due to limited experimental testing resources, and the remaining critical elements are numerically modelled along with the rest of the structure using the mathematical model assumptions that may not be suitable to predict the response of such critical elements \cite{eatherton2010large,fatemi2020experimental}. For instance, Imanpour et al. \cite{imanpour2018development} physically tested one of two critical columns of a steel multi-tiered concentrically braced frame, which may affect the overall response of the frame or even instability of the column tested in the laboratory. There are several other challenges associated with hybrid simulation, such as errors arising from test controller and hydraulic actuators, signal delays, and noise generated in instrumentation.

Machine-learning algorithms such as Prandtl neural network \cite{sharghi2019neuro}, and least-squares support-vector machine \cite{farrokh2018hysteresis} offer an efficient solution to some of the challenges described above. Such algorithms have been used in the past to develop meta-models for experimental substructures. Additionally, meta-models were created for finite element models in RTHS using the linear regression algorithm, and the recurrent neural network \cite{bas2020using}.

This paper proposes a new substructuring technique that employs a novel data-based model for the seismic response evaluation of steel braced frames, and presents preliminary results used for verification purposes. The proposed data-based model can learn the nonlinear hysteretic behaviour of structural elements (i.e., steel braces) using the sparse identification algorithm. This model is first trained using a pushover analysis performed on a numerical model of a steel buckling restrained brace (BRB), replacing actual experimental test data. A hybrid model of a prototype buckling-restrained braced frame (BRBF) is then developed with a trained model simulating the hysteretic response of the BRB. The adequacy of data-driven hybrid simulation technique is finally evaluated by comparing the results of a dynamic analysis obtained using the proposed technique with those predicted by a pure numerical model of the same BRBF.

\section{Data-driven Hybrid Simulation (DDHS) Technique}
\begin{figure}[tbp]
	\centering
	\includegraphics[scale=0.5]{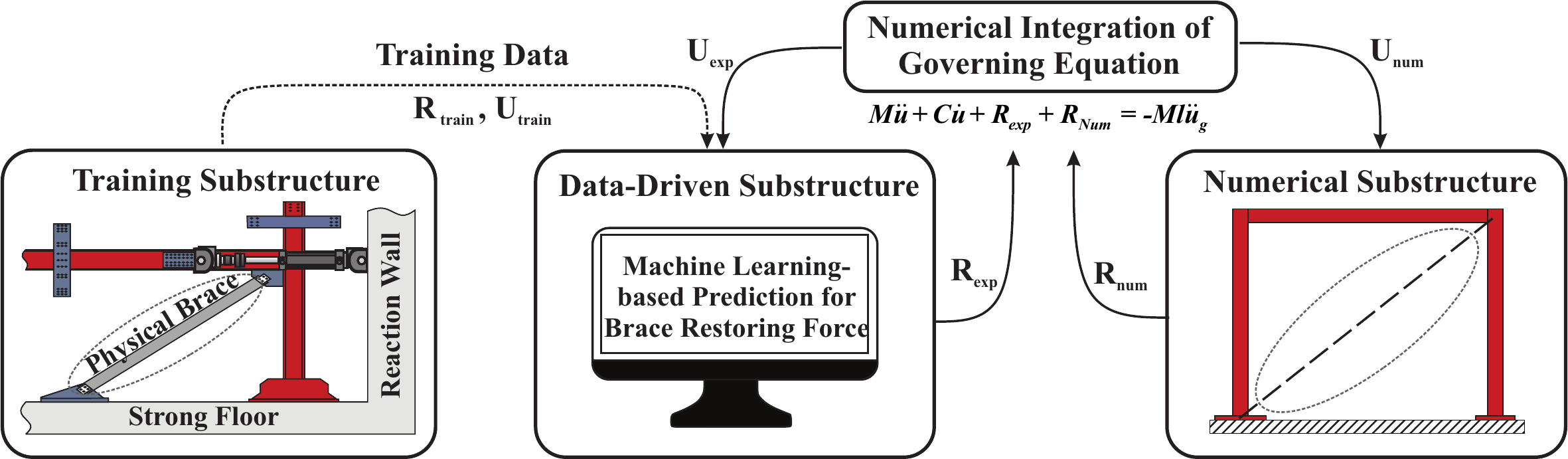}
	\caption{Concept of the data-driven hybrid simulation (DDHS) technique.} \label{Fig:DDHS}
	\squeezeup
\end{figure}

A new data-driven hybrid simulation technique is proposed here by incorporating two substructures, including a data-driven substructure that typically represents the structure's critical component and a finite element model that simulates the remaining components of the structure. The architecture of the proposed technique is shown in Fig. \ref{Fig:DDHS}. The data-driven substructure is powered by a machine learning algorithm trained using experimental test data capable of predicting the nonlinear dynamic response of critical components (e.g., seismic fuses in steel seismic force-resisting frames). The numerical substructure acting parallel to the data-driven substructure consists of validated elements that simulate the structural elements that essentially remain elastic (e.g., capacity-protected members in steel seismic force-resisting frames) under earthquake excitations.

\subsection{Sparse Identification of Nonlinear Dynamics (SINDy) Algorithm}
A new model discovery algorithm referred to as Sparse Identification of Nonlinear Dynamics (SINDy) was recently proposed by Brunton et al. \cite{brunton2016discovering} to extract the governing equation of dynamical systems from data using the sparse identification algorithm. SINDy will allow us to define the mathematical relationship, $R = f(x(t))$, between the component's force $R$ and the component's displacement $x(t)$ using the experimental data.
For this purpose, various potential functions are stacked in a library matrix, $\boldsymbol{\Theta}$:
\begin{equation}\label{eq:2}
	\boldsymbol{\Theta(\boldsymbol{X})}= \begin{bmatrix}
				\vline & \vline & \vline & \vline  \\
				g_1(\boldsymbol{X}) & g_2(\boldsymbol{X}) & \cdots  & g_m(\boldsymbol{X}) \\
				\vline & \vline & \vline & \vline 
			\end{bmatrix}_{n \times m}
\end{equation}
in which, $\boldsymbol{X}=[x(t_1),x(t_2),...,x(t_n)]^T$ is the displacement vector obtained from the experimental data used for training, $g_i(.), i=1,2,...,m$ are the candidate functions (i.e. basis functions), $n$ and $m$ are the number of measured data points and candidate functions selected to form $\boldsymbol{\Theta}$, respectively. The basis functions should be selected based on the nature of the phenomenon \cite{brunton2016discovering}. The higher number of basis functions involved in the library matrix can help improve discovering the relationship by the algorithm, but with more computational effort. By defining the library matrix, $\boldsymbol{\Theta}$, approximated restoring force vector of the component $\boldsymbol{\hat{R}}$ can be represented as:
\begin{equation} \label{eq:3}
	\underset{n \times 1}{\boldsymbol{\hat{R}}} = \underset{n \times m}{\boldsymbol{\Theta}} \times \underset{m \times 1}{\boldsymbol{\Xi}}
\end{equation}
in which, $\boldsymbol{\Xi}=[\xi_1,\xi_2,...,\xi_m]^T$ is the coefficient vector that determines which combination of the basis functions should be used to best approximate the nonlinearity involved in the problem. Sparse regression algorithms such as $L_1$-regularized regression (LASSO) can be employed to find the active entries of the $\boldsymbol{\Xi}$ vector. LASSO regression is chosen here due to its ability in shrinking irrelevant terms to zero, which helps promote sparsity and avoid overfitting \cite{tibshirani1996regression}. LASSO regression is formulated as the following minimization problem:
\begin{equation} \label{eq:4}
	minimize\; \frac{1}{2n} ||\boldsymbol{R}-\boldsymbol{\Theta \Xi}||_2^2 + \lambda ||\boldsymbol{\Xi}||_1
\end{equation}
in which, $\boldsymbol{R}$ is the restoring force vector obtained from training experimental data, $||.||_1$ and $||.||_2$ represent $L_1$-norm and $L_2$-norm, respectively. $\lambda$ denotes the regularization parameter that controls the sparsity of the solution. By solving the LASSO minimization problem, $\boldsymbol{\Xi}_{lasso}$ is obtained, which is then used to discover the the function, $R = f(x(t))$, as follows:
\begin{equation} \label{eq:5}
	\hat{R} = \boldsymbol{\Theta}(x) \times \boldsymbol{\Xi}_{lasso}
\end{equation}
where, $\boldsymbol{\Theta}(x)$ is the symbolic function of $x$ (displacement), in contrast to $\boldsymbol{\Theta}(\boldsymbol{X})$ that is the data library matrix of displacement defined in Eq. \ref{eq:2}.
\squeezeup
\subsection{Prandtl-Ishlinskii (PI) Model}
Given that the efficacy of SINDy highly depends on the basis functions selected in the library matrix, it is crucial to appropriately select such functions to well reproduce the hysteretic response of nonlinear systems. One method to complement SINDy with a hysteretic memory is achieved by utilizing hysteretic operators such as the stop operators in Prandtl-Ishlinskii (PI) model \cite{brokate1996hysteresis}.

PI model is a phenomenological operator-based model defined as the weighted superposition of multiple stop (elastic-perfectly plastic) operators. The stop operator, which was first introduced in the field of continuum mechanics to describe the elastoplastic behaviour of materials \cite{brokate1996hysteresis}, is recursively defined as follows:
\begin{equation} \label{eq:7}
	y_r(0) = e_r(x(0))
\end{equation}
\begin{equation} \label{eq:8}
	y_r(t) = e_r(x(t)-x(t_i)+y_r(t_i))\;\;\; for\; t_i < t \leq t_{i+1}; \;\; 0 \leq i \leq N-1
\end{equation}
\begin{equation} \label{eq:9}
	e_r(s)=min(r,max(-r,s))
\end{equation}
in which, $y_r(t)=E_r[x(t)]$ is the output of the stop operator for the given input signal $x(t)$, which is defined using a threshold $r$ $(r>0)$. Within each stop operator, the time domain $[0,T]$ needs to be discretized into $N$ subintervals of $0=t_0<t_1<...<t_N=T$ such that the input signal, $x(t)$, becomes monotonic within each time step $[t_i,t_{i+1}]$. By linearly combining multiple stop operators with different thresholds (r values), the discrete PI model can be defined:
\begin{equation} \label{eq:10}
	R(t) = \sum_{i=1}^{m}{\xi_i E_{r_{i}}[x(t)]}
\end{equation}
where, $R(t)$ is the total output signal when the hysteretic system is subjected to an exciting input signal ($x(t)$). The PI model parameters, the weights $\xi_i$ and thresholds $r_i$, should be optimized to fit the training data.
\squeezeup
\subsection{PI-SINDy Model for the Simulation of Hysteretic Response}
The proposed model, which is referred to as PI-SINDy, combines the PI model and the SINDy algorithm to simulate the nonlinear hysteretic response of structural components. In the PI-SINDy model, the stop operators play the role of basis functions of the library matrix (Eq. \ref{eq:2}), and the $\boldsymbol{\Xi}_{lasso}$ vector contains optimized weights ($\xi_i$) as given in Eq. \ref{eq:10} after the LASSO regression is carried out on the experimental data used for training. The PI-SINDy model is then defined as follows:
\begin{equation} \label{eq:11}
	\underset{n \times 1}{\boldsymbol{\hat{R}}} = \begin{bmatrix}
													\vline & \vline & \vline & \vline & \vline & \vline  \\
													E_{r_{1}}[\boldsymbol{X}] \;\; & E_{r_{2}}[\boldsymbol{X}] \;\; & \cdots \;\; & E_{r_{m}}[\boldsymbol{X}] \;\; & \boldsymbol{X} \;\; & \boldsymbol{1} \\
													\vline & \vline & \vline & \vline  & \vline & \vline
												  \end{bmatrix}_{n \times (m+2)} 
											  		\times 
											  		\underset{(m+2) \times 1}{\boldsymbol{\Xi}_{lasso}}
\end{equation}
In the proposed model, several potential thresholds $r_i$ need to be assumed for stop operators. Such thresholds should then be included in the library matrix so that SINDy can determine which combination of thresholds can best describe the hysteretic response of the experimental data used for training. The initial set of thresholds are defined based on the maximum deformation ($|x_{max}|$) as $r_i = \frac{i}{m+1} |x_{max}|, \; i = 1,2,...,m$, in which, $m$ can be obtained using a sensitivity analysis. It is worth noting that the larger value of $m$ would result in a finer discretization but a computationally less-effective optimization.

\section{Numerical Verification} \label{sec:NumVer}
\begin{figure}[t]
	\centering
	\includegraphics[scale=0.5]{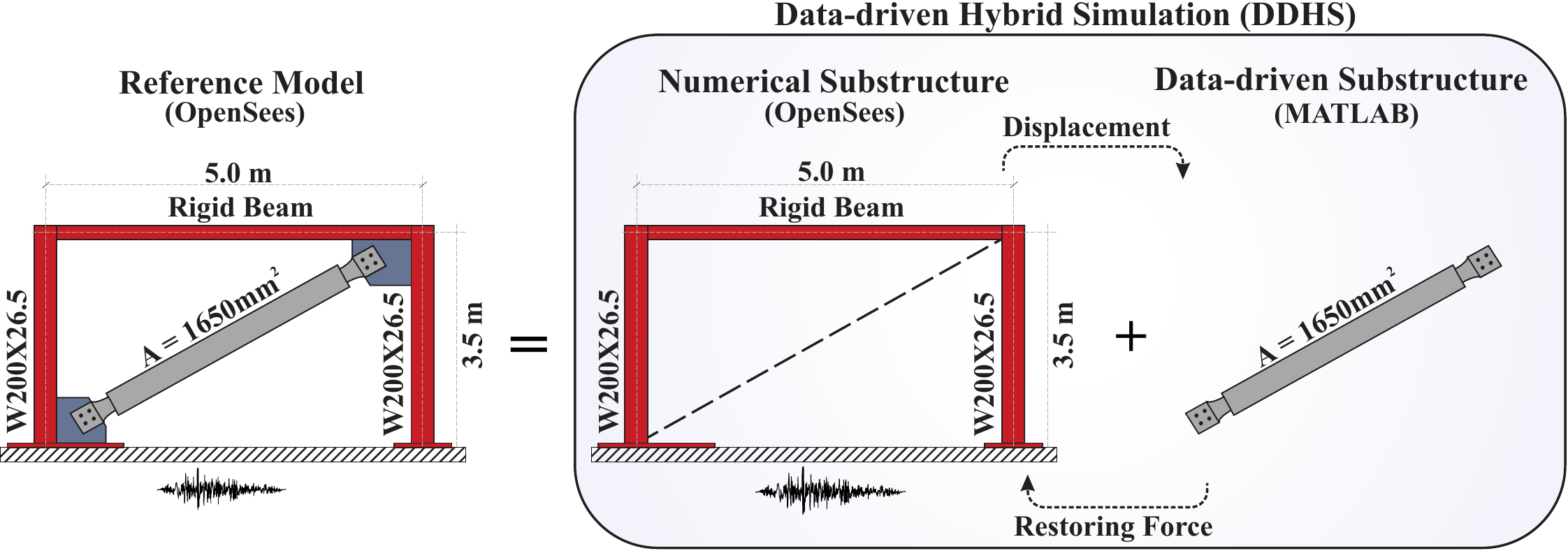}
	\caption{Simulation of the prototype buckling-restrained braced frame using pure numerical model and the proposed data-driven hybrid simulation technique.} \label{Fig:NumVer}
	\squeezeup
\end{figure}

The efficiency of DDHS was evaluated using the nonlinear response history analysis (NLRHA) of a prototype two-dimensional (2-D) buckling-restrained braced frame (BRBF). The properties of the frame are shown in Fig. \ref{Fig:NumVer}. The pure numerical model of the BRBF is developed in the \textit{OpenSees} program \cite{mckenna1999object}. The beam and columns were modelled using elastic beam-column elements, and the Giuffré-Menegotto-Pinto material model assigned to a truss element was used to construct the BRB \cite{filippou1983effects}. For simplicity, the influence of frictional forces developed between the BRB core and restraint was neglected in the model, resulting in a symmetric hysteretic response under cyclic loading in tension and compression. The fundamental period of this structure is 0.492 seconds.

\subsection{Training of the BRB using the PI-SINDy model} \label{sec:NumVer_Training}
\begin{figure}[t]
	\centering
	\includegraphics[scale=0.5]{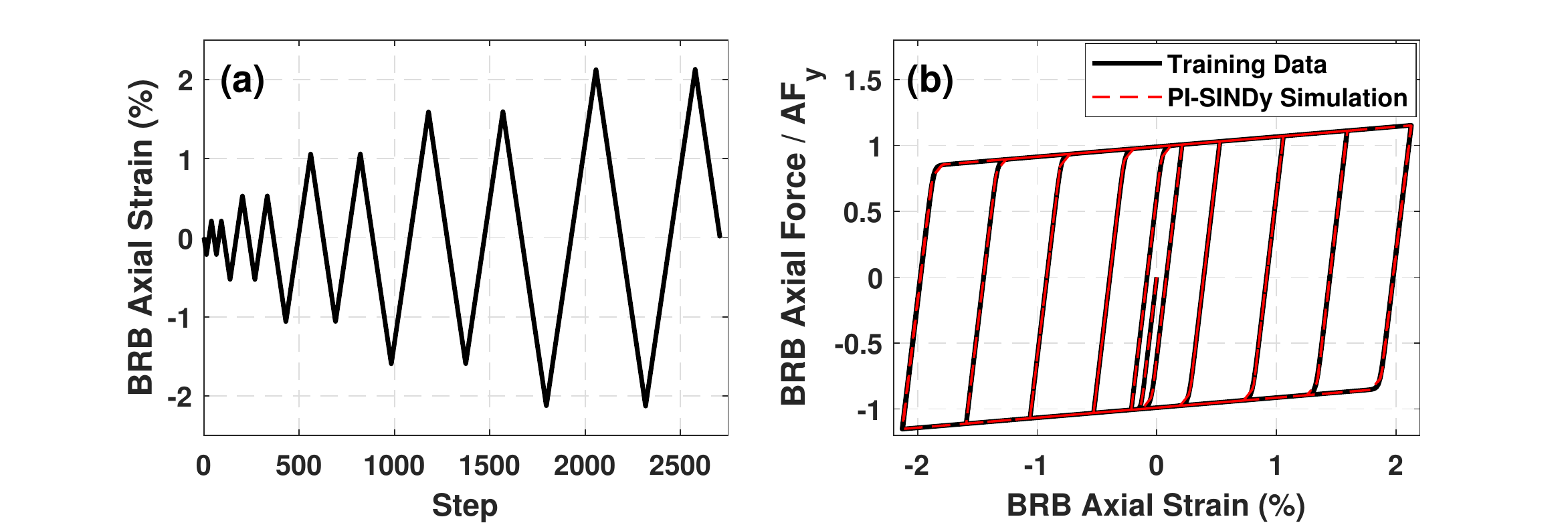}
	\caption{PI-SINDy training: (a) Displacement protocol, (b) Training data generated by a pushover analysis on the BRB and PI-SINDy simulation after training.} 
	\label{Fig:Train}
	\squeezeup
\end{figure}

In lieu of actual experimental test data, fictitious test data is used in training, which was created using a cyclic nonlinear static (pushover) analysis performed on an identical isolated BRB under an increasing cyclic displacement protocol shown in Fig. \ref{Fig:Train}a. PI-SINDy model was then trained with this input as described in Algorithm \ref{algo:PISINDy}. 50 stop operators were selected to be used in the data library matrix, and $\lambda$ was taken as 0.1 in LASSO regression. Normalized root-mean-square-error (NRMSE) as given in Eq. \ref{eq:13} was used to compute the error between the approximated hysteretic response by the PI-SINDy model and that obtained from the pure numerical model.
\begin{equation} \label{eq:13}
	NRMSE = \frac{\sqrt{\sum_{i=1}^{N}(y_{ref,i}-y_{model,i})^2/N}}{|y_{ref,max}-y_{ref,min}|}
\end{equation}
Fig. \ref{Fig:Train}b shows the cyclic response of the isolated BRB obtained from the cyclic pushover analysis, compared with the BRB response simulated by PI-SINDy. As shown, the PI-SINDy successfully captured the nonlinear cyclic behaviour of the BRB with NRMSE = $0.1954\%$.
\begin{algorithm}[b]
	\SetAlgoLined
	\SetKwInOut{Input}{input}
	\SetKwInOut{Output}{output}
	\Input{
		(i) Experimental deformation history: $\boldsymbol{X}=[x(t_1),x(t_2),...,x(t_n)]^T$ \newline
		(ii) Experimental restoring force: $\boldsymbol{R}=[R(t_1),R(t_2),...,R(t_n)]^T$}
	Calculate thresholds $r_i = \frac{i}{m+1} |x_{max}| \;\; i = 1,2,...,m$ \\
	\For{$i=1:m$}{
		Calculate $E_{r_{i}}[\boldsymbol{X}]$: Eqs. \ref{eq:7}, \ref{eq:8}, \ref{eq:9} \\
		$\boldsymbol{\Theta}(:,i)=E_{r_{i}}[\boldsymbol{X}]$: Eq. \ref{eq:11}
	}
	LASSO regression: $minimize\; \frac{1}{2n} ||\boldsymbol{R}-\boldsymbol{\Theta \Xi}||_2^2 + \lambda ||\boldsymbol{\Xi}||_1 	\rightarrow find\; \boldsymbol{\Xi}_{lasso}$ in Eq. \ref{eq:11}
	
	\caption{PI-SINDy Algorithm}
	\label{algo:PISINDy}
\end{algorithm}

\subsection{Verification of the DDHS Technique}

The BRB in the BRBF of Fig. \ref{Fig:NumVer} was replaced with a PI-SINDy-powered data-driven substructure to verify the capability of the proposed DDHS technique in predicting the restoring force of the BRB when the frame is subjected to dynamic loading. The training process of the PI-SINDy model is described in Section \ref{sec:NumVer_Training}. PI-SINDy developed in MATLAB \cite{MATLAB:R2019b} was linked to the numerical substructure in \textit{OpenSees}. The UT-SIM framework \cite{huang2020generalized,UTSIMmanual} was used to establish communication between two substructures. An NLRHA \cite{auger2016multi} was performed under the 2011 Miyagi MYG016 record applied in the horizontal direction of the frame in the \textit{OpenSees} program. At the analysis time step, the displacement was sent to MATLAB to predict the restoring force. The predicted force vector in interfacial DOFs was then fed back to \textit{OpenSees} to complete the numerical integration of the equation of motion at that time step.

\begin{figure}[t]
	\centering
	\makebox[\textwidth][c]{\includegraphics[width=1.15\textwidth]{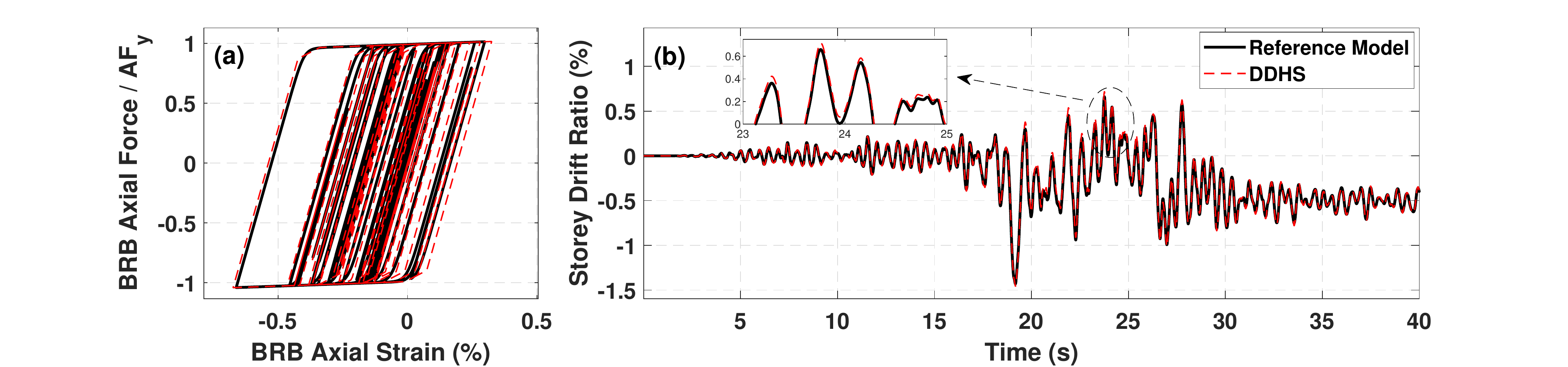}}
	\caption{Results of nonlinear response history analysis for the prototype frame: (a) BRB hysteretic response, and (b) History of the storey drift ratio (for the first 40 s of the ground motion duration shown).}
	\label{Fig:Result}
\end{figure}

Fig. \ref{Fig:Result} portrays the BRB hysteretic response and the history of storey drift ratio obtained from the NLRHA of the DDHS technique and the reference model. As shown, BRB responded in the inelastic range and showed a symmetrical response as expected. The frame underwent a maximum drift ratio of 1.43\%, while the BRB experienced a maximum axial strain equal to 0.66\%. The NRMSE of the storey drift ratio is 0.7702\%, which implies the accuracy of the DDHS technique using the trained symmetrical hysteresis response of the BRB.

\section{Conclusions}
This paper proposes a new data-driven substructuring technique for the seismic evaluation of steel braced frame structures that is accomplished through incorporating a novel data-driven model (PI-SINDy) predicting the nonlinear cyclic response of the frame’s critical elements combined with a fibre-based numerical substructure simulating the well-understood components of the frame. PI-SINDy was developed using the sparse identification algorithm and inspired by the Prandtl-Ishlinskii model to predict the hysteretic response of BRB using available experimental test data. The preliminary results confirm that once PI-SINDy is trained, it can properly emulate the nonlinear cyclic response of a physical substructure. Moreover, for a symmetric hysteresis response, cyclic test data obtained from quasi-static testing would be enough to train PI-SINDy and reproduce the brace hysteretic response under random signals such as earthquake accelerations. The numerical verification results on the prototype single storey frame show the feasibility and suitability of the PI-SINDy model. Future studies should be devoted to further improve the proposed data-driven model for more complex hysteretic responses typically observed in steel braced frames. Furthermore, verification tests should be carried out in the future to verify the application of the proposed DDHS technique.

\section{Acknowledgments}
Funding for this project has been provided by the National Sciences and Engineering Research Council (NSERC). The authors gratefully acknowledge the CISC Centre for Steel Structures Education and Research at the University of Alberta for their continuous support, and the UT-SIM framework team at the University of Toronto for their assistance in the development of the hybrid simulation platform. The first author acknowledges scholarships from the government of Alberta, the Norman and Tess Reid Family, and the Parya foundation. 
\squeezeup
\bibliographystyle{splncs04}
\bibliography{References}

\begin{thebibliography}{10}
\providecommand{\url}[1]{\texttt{#1}}
\providecommand{\urlprefix}{URL }
\providecommand{\doi}[1]{https://doi.org/#1}

\bibitem{auger2016multi}
Auger, K., Minouei, Y., Elkady, A.M.A., Imanpour, A., Leclerc, M., Lignos, D.,
  Toutant, G., Tremblay, R.: Multi-directional structural component hybrid
  testing system for the assessment of the seismic response of steel i-shaped
  columns. In: Proceedings of the 11th Pacific Structural Steel Conference.
  No.~CONF (2016)

\bibitem{bas2020using}
Bas, E.E., Moustafa, M.A., Feil-Seifer, D., Blankenburg, J.: Using machine
  learning approach for computational substructure in real-time hybrid
  simulation. arXiv preprint arXiv:2004.02037  (2020)

\bibitem{brokate1996hysteresis}
Brokate, M., Sprekels, J.: Hysteresis and phase transitions, vol.~121. Springer
  Science \& Business Media (1996)

\bibitem{brunton2016discovering}
Brunton, S.L., Proctor, J.L., Kutz, J.N.: Discovering governing equations from
  data by sparse identification of nonlinear dynamical systems. Proceedings of
  the national academy of sciences  \textbf{113}(15),  3932--3937 (2016)

\bibitem{eatherton2010large}
Eatherton, M.R.: Large-scale cyclic and hybrid simulation testing and
  development of a controlled-rocking steel building system with replaceable
  fuses. University of Illinois at Urbana-Champaign (2010)

\bibitem{farrokh2018hysteresis}
Farrokh, M.: Hysteresis simulation using least-squares support vector machine.
  Journal of Engineering Mechanics  \textbf{144}(9),  04018084 (2018)

\bibitem{fatemi2020experimental}
Fatemi, H., Paultre, P., Lamarche, C.P.: Experimental evaluation of inelastic
  higher-mode effects on the seismic behavior of rc structural walls. Journal
  of Structural Engineering  \textbf{146}(4),  04020016 (2020)

\bibitem{filippou1983effects}
Filippou, F.C., Popov, E.P., Bertero, V.V.: Effects of bond deterioration on
  hysteretic behavior of reinforced concrete joints  (1983)

\bibitem{huang2020generalized}
Huang, X., Kwon, O.S.: A generalized numerical/experimental distributed
  simulation framework. Journal of Earthquake Engineering  \textbf{24}(4),
  682--703 (2020)

\bibitem{imanpour2018development}
Imanpour, A., Tremblay, R., Leclerc, M., Siguier, R.: Development of a hybrid
  simulation computational model for steel braced frames. In: Key Engineering
  Materials. vol.~763, pp. 609--618. Trans Tech Publ (2018)

\bibitem{MATLAB:R2019b}
The Mathworks, Inc., Natick, Massachusetts: {MATLAB version 9.7.0.1216025
  (R2019b) Update 1} (2019)

\bibitem{mccrum2016overview}
McCrum, D., Williams, M.: An overview of seismic hybrid testing of engineering
  structures. Engineering Structures  \textbf{118},  240--261 (2016)

\bibitem{mckenna1999object}
Mckenna, F.T.: Object-oriented finite element programming: Frameworks for
  analysis, algorithms and parallel computing.  (1999)

\bibitem{UTSIMmanual}
Mortazavi, P., Huang, X., Kwon, O.S., Christopoulos, C.: Example manual for the
  university of toronto simulation framework. an open-source framework for
  integrated multi-platform simulations for structural resilience (second
  edition) (07 2017)

\bibitem{sharghi2019neuro}
Sharghi, A.H., Karami~Mohammadi, R., Farrokh, M.: Neuro-hybrid simulation of
  non-linear frames using prandtl neural networks. Proceedings of the
  Institution of Civil Engineers-Structures and Buildings pp. 1--18 (2019)

\bibitem{tibshirani1996regression}
Tibshirani, R.: Regression shrinkage and selection via the lasso. Journal of
  the Royal Statistical Society: Series B (Methodological)  \textbf{58}(1),
  267--288 (1996)

\end{thebibliography}

\end{document}